# Development and Evaluation of Video Recordings for the OLSA Matrix Sentence Test

Gerard Llorach[†,1,2,3]
gerard.llorach.to@uni-oldenburg.de

Frederike Kirschner[2,3]
frederike.kirschner@uni-oldenburg.de

Giso Grimm[2,3]
g.grimm@uni-oldenburg.de

Melanie A. Zokoll[1,2]
m.zokoll@hoerzentrum-oldenburg.de

Kirsten C. Wagener[1,2,4]
k.wagener@hoerzentrum-oldenburg.de

Volker Hohmann[1,2,3]
volker.hohmann@uni-oldenburg.de

[1] Hörzentrum Oldenburg GmbH
Oldenburg, Germany

[4] Hörtech gGmbH
Oldenburg, Germany

[2] Cluster of Excellence Hearing4All
Dept. of Medical Physics and Acoustics
University of Oldenburg
Oldenburg, Germany

[3] Auditory Signal Processing
Dept. of Medical Physics and Acoustics
University of Oldenburg
Oldenburg, Germany

**ABSTRACT**

**Objective:** The aim was to create and validate an audiovisual version of the German Matrix Sentence Test, which uses the existing audio-only speech material.

**Design:** Video recordings were recorded and dubbed with the audio of the existing German matrix sentence test (MST). The current study evaluates the MST in conditions including audio and visual modalities, speech in quiet and noise, and open and closed-set response formats.

**Sample:** 1 female talker recorded repetitions of the German MST sentences. 28 young normal-hearing participants completed the evaluation study.

**Results:** The audiovisual benefit in quiet was 7 dB in sound pressure level (SPL). In noise, the audiovisual benefit was 4.9 dB in signal-to-noise ratio (SNR). Speechreading scores ranged from 0% to 84% speech reception in visual-only sentences (mean = 50%). Audiovisual speech reception thresholds (SRTs) had a larger standard deviation than audio-only SRTs. Audiovisual SRTs improved successively with increasing number of lists performed. The final video recordings are openly available.

**Conclusions:** The video material achieved similar results as the literature in terms of gross speech intelligibility, despite the inherent asynchronies of dubbing. Due to ceiling effects, adaptive procedures targeting 80% intelligibility should be used. At least one or two training lists should be performed.

† Corresponding author

## 1 Introduction

Speech audiometry is an essential element in audiology [1][2]. It assesses the ability to understand speech acoustically, which is crucial for human communication. The matrix sentence test (MST) [45] is a well-established method in speech audiometry, and it exists in several languages [3]. MSTs use sentences of 5 words with a "noun - verb - number - adjective - object" structure. There are 10 possible words for each word category (e.g., 10 nouns, 10 verbs, etc.); these are combined to create semantically unpredictable, syntactically correct sentences. Lists of 20 sentences are commonly used to test speech intelligibility.

Although speech can be understood through sounds only, it is a multimodal process. Being able to see the speaker provides additional cues such as lip movements, which make speech much easier to understand [4]. Audiovisual speech perception has been mentioned as a predictor of real-world hearing disability [5] but it is usually not considered in audiometry [42]. Visual information supports speech intelligibility, particularly severely impaired listeners are relying on visual information in adverse listening conditions [10]. The MST is also intended as a speech test for severely impaired listeners, therefore an audiovisual version is an important extension for its applicability. Nevertheless, audiovisual (or auditory-visual) MSTs with video recordings have only been developed in Malay, New Zealander English, and Dutch [6][7][8].

The ability to speechread (most commonly known as lipreading) plays a key role in audiovisual speech tests. In particular, audiovisual MSTs are highly affected by speechreading ability. In the Malay MST [6] young, normal-hearing participants scored from 25% to 85% speech reception just by speechreading, i.e., in the visual-only condition. Such visual-only scores indicate that participants are able to understand speech without any acoustic cues. This means that there is a ceiling effect in the audiovisual MSTs: even if speech is completely masked by noise and not heard, participants achieve their visual-only score.

Recording and validating an MST is quite an extensive undertaking: selection of the phonetically balanced speech material, recording of the speech, cutting and processing of the sound files, making each word equally intelligible to the others, evaluation, and validation [3].

In order to reduce cost and effort in the creation of an audiovisual MST from scratch, existing audio-only MST can be reused. Because audio-only MSTs already exist and have been used extensively, it is reasonable to reuse the audio material in audiovisual tests. New audio recordings cannot be compared directly to other recordings of the same language, as the speaker influences the intelligibility of the MST (up to 6 dB differences between talkers) [44]. Reusing the audio material ensures validity across studies, and saves time and effort. If the audio recordings are newly created, they need to be optimized to allow for a steep intelligibility function (a prerequisite for an accurate test), which includes measuring the intelligibility functions for each word of the test in a large number of participants. This would multiply the effort in comparison to producing dubbed videos.

One approach that has been proposed uses virtual characters with lip-synchronization together with existing audio-only speech tests [9, 10, 39]. The advantage of virtual characters is that they can be set in different configurations with relatively little effort [11]. The proposed approach in this paper is to create video recordings dubbed with existing audio for speech tests. A video recording usually provides better quality and realism than a virtual character. Nevertheless, asynchronies between the audio and the video have to be kept below 45ms (audio ahead) and 200ms (audio delayed) in order to pass unnoticed [12] and not affect speech intelligibility [13]. Additionally, further considerations must be taken into account, such as the head movements and facial expressions of the speaker [6].

One of the advantages of MSTs is that the sentences are unpredictable and there are too many word combinations to be memorized, so consecutive tests in different conditions can be carried out. Nevertheless, the simple sentence structure and the limited number of words enable participants to learn and improve their results. This training effect has already been shown in audio-only MSTs [14][15] and is particularly noticeable in the first list of 20 sentences, where differences in SRTs of about 1 dB are expected. After 2-4 lists, there is usually an absolute improvement of 2 dB, and the training effects in the following lists are quite small. In audiovisual MSTs, it is expected that participants further improve their SRTs by becoming familiar with the speaker and the visual material [16] and because training effects have been found to be stronger in audiovisual speech [41].

Another factor to take into account is the response format of the MST. After hearing a sentence, participants either repeat what they heard (open-set response format) or select the answers from all possible words (closed-set response format). In the open-set format, a researcher must be present in order to assess whether the answer is correct, while in the closed-set format, participants can do the test by themselves. The closed-set format may give participants an advantage, since they are provided with a list of all possible words; in fact, SRTs have been found to be lower with closed-set type in some MSTs [17][18], although not for German and other languages [3]. Whether such effects appear in audiovisual MSTs has not yet been investigated.

In this work we created an audiovisual version of the female German MST (AV-OLSAf). We recorded videos with a female speaker, dubbed them with the original sentences of the female speaker [15][19] and evaluated the material. Our first contribution is the methodology for producing the dubbed videos and getting the best synchronized video recordings. The final video recordings for the AV-OLSAf can be found in [40]. Our second contribution is the evaluation of the AV-OLSAf with normal-hearing listeners in different conditions: we show the audiovisual training effects in the open-set and closed-set responses; we discuss the speechreading scores and the effects of speechreading in the audiovisual SRTs; and we compare the audio-only and audiovisual SRTs in noise and in quiet conditions. To conclude, we discuss the implications and recommendations for using the AV-OLSAf.

## 2 Method

### 2.1 Recording the Video Material

Although in theory there are 100,000 possible sentences (5 word categories with 10 words per category, Table 1), the female OLSA uses only 150 predetermined sentences. This relatively small number of sentences permitted us to record videos of the spoken sentences in a single afternoon. We were able to recruit the same speaker that recorded the audio-only version of the German female MST (OLSA) [15][19]. She was a speech therapist and a singer. During the recording session, the speaker had to speak the sentences simultaneously while hearing them through an earphone on the right ear. Each sentence was played five times consecutively. Three short "beep" signals were given before each repetition started. The first repetition was used as a reference: the speaker was to listen only in order to know what sentence was coming. In the remaining 4 repetitions, she was to speak simultaneously while hearing the sentence.

**Table 1. Set of words used in the German Matrix Sentence Test. The sentences are combinations of 5 words from different categories, e.g., "*Doris malt neun nasse Sessel*" or

"<u>Nina</u> <u>bekommt</u> <u>vier</u> <u>rote</u> <u>Schuhe</u>". **The order shown here is the same as it was shown to the participants in the closed-set response format.**

| Noun | Verb | Number | Adjective | Object |
|---|---|---|---|---|
| Britta | <u>bekommt</u> | zwei | alte | Autos |
| *Doris* | gewann | drei | große | Bilder |
| Kerstin | gibt | <u>vier</u> | grüne | Blumen |
| <u>Nina</u> | hat | fünf | kleine | Dosen |
| Peter | kauft | sieben | *nasse* | Messer |
| Stefan | *malt* | acht | <u>rote</u> | Ringe |
| Tanja | nahm | *neun* | schöne | <u>Schuhe</u> |
| Thomas | schenkt | elf | schwere | *Sessel* |
| Ulrich | sieht | zwölf | teure | Steine |
| Wolfgang | verleiht | achtzehn | weiße | Tassen |

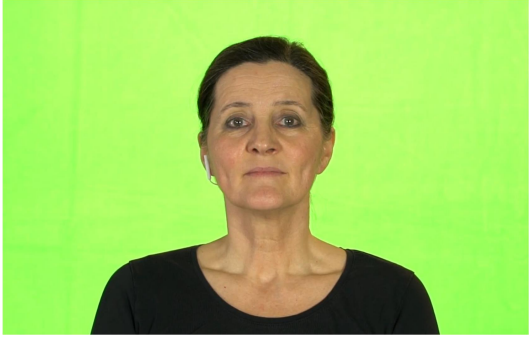

**Figure 1: Example of a frame of the video material.**

The videos of the female speaker were recorded in the studio of the Media Technology and Production of the CvO University of Oldenburg. The available lights of the studio were set up to achieve a homogeneous illumination of the face and of the background green chroma key. The videos were recorded with a Sony α7S II camera at 50pfs / full HD, and a condenser microphone in front of the speaker at the height of the knees. The speech was recorded in one channel with a 48 kHz sampling rate and a 16 bit linear pulse-code modulation (LPCM) sample format. An image sample of the final video recordings is shown in Figure 1.

A computer was used to reproduce the original OLSA sentences, which at the same time was sending a linear time code (LTC) signal to the second audio channel of the camera. This way, the recorded speech of the session and the original sentences could be synchronized. The recording session lasted around 2 hours in total.

## 2.2 Selection of the Videos

We manually discarded videos in which the speaker smiled or showed other non-neutral facial expressions. The recorded speech signals were synchronized to the reproduced original sentences using the LTC signal. When dubbing speech, there are inevitable asynchronies: time offsets (words spoken too early or too late) and/or words spoken slower or faster than the original words. As all these asynchronies could happen in one single sentence, we used dynamic time warping (DTW) [20] to find the best match between the recordings and the original sentences. The DTW quantified the temporal misalignment between the original and recorded sentences. The algorithm compares two temporal signals and provides a warping path. We computed the mel spectrograms of the signals and used them for the DTW function. The mel spectrograms were done using frame windows of 46 ms with a frame shift of 23 ms. An example of the mel spectrograms and the corresponding warping path can be seen in Figure 2 and Figure 3. Once the warping path was calculated, we used equations 1 and 2 to compute the asynchrony score:

$$wp_{ij}(n) = DTW(melSpec_i(n), melSpec_j(m)) \quad (1)$$

$$async_{ij} = RMS(wp_{ij}(n) - n) \quad (2)$$

for $i = 1,2,3,...,150$ (original sentence number) and
$j = 1,2,3,4$ (recording nº per original sentence)

where the $melSpec_i(n)$ is the mel spectrogram of the original sentence $i$, $melSpec_j(n)$ is the mel spectrogram of its corresponding recording (4 recordings $j$ per sentence $i$), $n$ is the frame number of the $melSpec_i$, $m$ is the frame number of the $melSpec_j$, $wp_{ij}(n)$ is the warping path between the mel spectrograms in frames, $(wp_{ij}(n) - n)$ is the difference in frames, $RMS$ is the root mean square, and $async_{ij}$ is the asynchrony score between the $i^{th}$ original sentence and the $j^{th}$ recording of that sentence. The RMS was used because it represents the asynchrony score over a whole sentence. As our main interest is the speech intelligibility of the whole sentence, we did not consider momentary asynchronies, such as maximum asynchrony, as a measure to choose the best video recording. The asynchrony score can be further expressed in seconds instead of frames, as it represents a temporal difference.

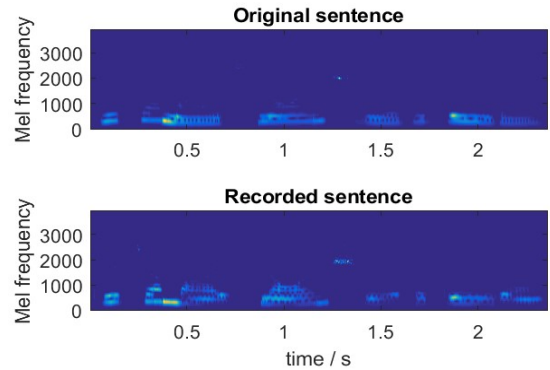

**Figure 2: Mel spectrogram of original sentence and one of the four recordings of that sentence.**

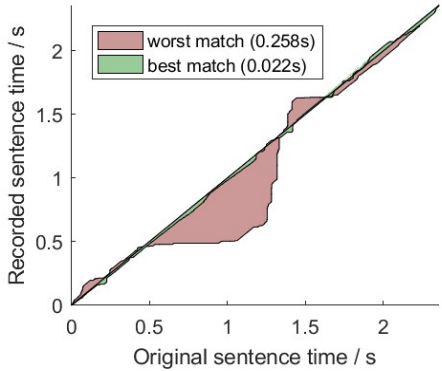

**Figure 3:** Warping path between the original and two recorded sentences. The best match and the worst match are shown. The size of the shaded surface corresponds to the asynchrony score.

We checked the sensitivity of this measure by comparing each recording to its corresponding original sentence and to the remaining unmatching original sentences (Figure 4). For each original sentence, we chose the video recording with the smallest asynchrony score. Of the final best selections, we found three outliers, with asynchrony scores greater than 80 ms, that had to be manually corrected with time offsets. Once corrected, these outliers were shown to 5 normal-hearing participants along with the best-matched sentences; the outliers could not be distinguished from the best-matched sentences and no asynchronies were noticed. We decided that the mean asynchrony score was small enough (~40 ms) to minimize the perceptual asynchrony/dubbing effects when measuring speech intelligibility with lists of 20 sentences: in [13], the authors evaluated the speech intelligibility of different timing misalignments with video and audio. According to them, visual asynchronies from -45ms to +200ms are not perceivable and speech recognition does not decline. Therefore, we proceeded with the evaluation of the material. The asynchrony score, maximum asynchronies and asynchrony over time of each sentence can be found in the supplemental material. The final video recordings can be found in [40].

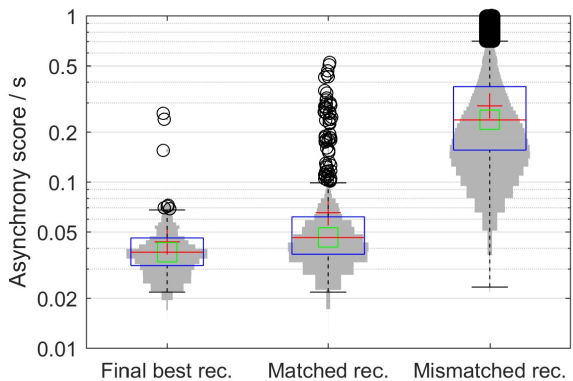

**Figure 4:** Asynchrony scores comparing the original sentences and their best-matched recordings before manual correction of the three outliers (left; 150 scores), the original sentences and all 4 of their recordings (middle; 600 scores) and the original sentences and the mismatched recordings (right; 150x149x4 scores). The vertical axis is on a logarithmic scale. The mean is represented as a red cross and the median as a green square. The outliers are depicted with black circles.

## 2.3 Evaluation of the Audiovisual Material

*2.3.1 Participants.* 28 normal-hearing participants (14 female, 14 male) took part in the evaluation measurements. Their ages ranged from 20 to 29 years (mean age 24.9 years). They had normal or corrected-to-normal vision and their pure tone averages (PTAs) in the better ear were between -5 and 7.5 dB HL (mean -0.31 dB HL). The PTAs were computed using the frequencies 0.5, 1, 2 and 4 kHz. Participants were recruited through the database of the Hörzentrum Oldenburg GmbH and were paid an expense allowance. Permission was granted by the ethics committee of the CvO University of Oldenburg.

*2.3.2 Setup.* Participants were seated in a chair inside a soundproof booth. The evaluation measurements were done using binaural headphones (Sennheiser HDA 200). A 22" touchscreen display with full HD (ViewSonic TD2220, ViewSonic Corp. Walnut, CA, USA) was placed in front of the participant within arm's reach at a height of 0.8 meters. The experiment was programmed in Matlab2016b. The videos and original sentences were reproduced with VLC 3.03. The acoustic signal was routed with RME Total Mix with an RME Fireface 400 sound card.

The acoustic levels were calibrated using a sound level meter placed at the approximate head position where participants would be seated. The sound and video reproduction was calibrated for synchronization using an external camera. For this purpose, we reproduced a video with frame numbering together with a LTC signal using the experiment setup. The external camera recorded the display screen of the experiment. The LTC signal was connected directly to the external camera instead of the headphones. Using the recording of the external camera we found a consistent asynchrony of 80 ms, which we corrected by delaying the audio signal in the experiment setup.

*2.3.3 Stimuli.* The acoustic stimulus was the female version of the German matrix sentence test (OLSA) [15] [19] and the visual stimuli was the best-matched video recording (see Section 2.2). For the conditions with noise, we used continuous test-specific noise (TSN) based on the female speech material. The presentation level of the noise was kept constant at 65 dB SPL. The speech level of the first sentence was 60 dB SPL for conditions with and without noise. The adaptive procedure used varied the speech presentation level depending on the responses of the participant.

*2.3.4 Conditions.* There were nine conditions in the experiment (see Table 2). Each condition used a list of 20 sentences. The sentences in each list were predefined by the MST. In total, we used 45 different predefined lists. The speech presentation levels were adapted after each sentence in order to reach an individual SRT of 80%, i.e. 4 out of 5 words correctly recognized per sentence. During the open-set response format, participants were asked to repeat orally what they understood after each sentence. In the closed-set response format, participants chose the words they understood from an interface displayed on the touch screen after stimulus presentation. The closed-set interface showed all 50 possible words plus one no-answer option per word category. In the visual-only condition (VONoiseClosed), there was no acoustic speech but only test-specific noise at 65 dB SPL. In this condition, the speech could only be understood through speechreading. For this condition, the percentage of correct words per sentence was averaged over 20 sentences (a list). In all conditions, no feedback was given about correctness of responses.

*2.3.5 Adaptive procedure.* We chose a SRT of 80% to avoid ceiling effects in audiovisual conditions due to the visual-only contribution, i.e., some participants might be able to understand more than 50% of the content just by speechreading [6][7][8]. The adaptive procedure used in this experiment is described in [21] and in [43]. It is an extended staircase method that changes its step size depending on the responses. The change in the presentation level is done in two stages. The first stage follows the equation presented in [21]:

$$\Delta L = -\frac{f(i) \cdot (prev - tar)}{slope} \quad (1)$$

where $\Delta L$ is the increment level, *prev* is the current result, *tar* is the target value, and *slope* is set to 0.1 dB$^{-1}$ in this study. The function *f(i)* defines the convergence rate, where *i* is the number of reversals in the presentation level, i.e. *i* increases every time the participant goes from being above/below threshold. In our study the current result is the discrimination value of the previous sentence and the target value is 0.8 (80% SRT). The value of *f(i)* is defined by 1.5 / 1.41$^i$ and its set to 0.25 for $i \leq 6$. The step size gets smaller when the participant crosses the target value. The second stage is described and examined in [43]. In this second stage, the step size is multiplied by 2 when two conditions are met: the step is a decrement (it lowers the presentation level) and *f(i)* is bigger than 0.5. This last condition is usually met in the first sentences of a list.

$$\Delta L = \begin{cases} 2 \cdot \Delta L & for\ f(i) \geq 0.5\ and\ \Delta L < 0 \\ \Delta L & otherwise \end{cases} \quad (2)$$

The final level estimate of a list is computed using a maximum-likelihood method and discrimination function described in [21].

**Table 2. Conditions tested for the evaluation and validation of the AV-OLSA.**

| | | Audio-only (AO) | Audiovisual (AV) | Visual-only (VO) |
|---|---|---|---|---|
| Noise | Closed-set response | **AONoiseClosed** | **AVNoiseClosed** | **VONoiseClosed** |
| | Open-set response | **AONoiseOpen** | **AVNoiseOpen** | |
| Quiet | Closed-set response | **AOQuietClosed** | **AVQuietClosed** | - |
| | Open-set response | **AOQuietOpen** | **AVQuietOpen** | |

*2.3.6 Training.* We added training lists prior to evaluating the nine conditions in order to assess the training effects of the AV-OLSA. We also tested the participants in two different sessions (test, retest). In the first session, 4 audiovisual lists were presented in noise (80 sentences in total). Participants were randomly assigned to do the 4 training lists in open-set or closed-set formats (AVNoiseClosed or AVNoiseOpen); 13 participants completed the training in the closed-set format and 15 participants in the open-set format. In the second session, the training was a single list with the same format as the first session (20 sentences in AVNoiseClosed or AVNoiseOpen).

*2.3.7 Procedure.* For the test lists, participants started with the same response format (open-set or closed-set) as in the training. Next, they did the conditions with the opposite response format. The conditions of one response format were presented in pseudo-randomized order (Figure 5). On the retest session, participants performed a training list with the same response format as on training lists of the first session; then they continued with the conditions with that same format before doing the ones with the other format, as on the test session. The test and retest sessions were temporally spaced from one day to two weeks.

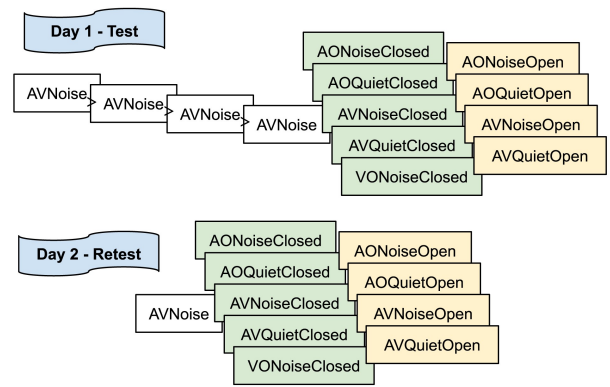

**Figure 5: Ordering of the lists in the test and retest sessions. Conditions stacked in columns (green and yellow) were pseudo-randomized within the column. If the participants**

were trained in AVNoise with the open-set format, they performed the open-set format lists before the closed-set lists; if they were trained with the closed-set format, they proceeded with the closed-set format lists before doing the open-set lists.

## 3 Results

For each list, a final level estimate was computed as described in Section 2.3.5. This value, i.e. the SRT at 80%, was expressed in dB SNR for the conditions in noise and in dB SPL for the conditions in quiet. For the VONoiseClosed lists it was different: the percentage of words understood over all 20 sentences was computed (i.e. the speechreading score). For each participant there were 5 audiovisual training lists, 4 in the first session and 1 in the second session, and 18 test lists, 9 in each session (see Figure 5). For the analysis of the results, we removed an outlier of +9 dB SNR belonging to an AONoiseClosed list of the first session (test).

### 3.1 Training Effects

In general, audiovisual SRTs tended to improve across lists. Figure 6 shows the mean SRTs during the training lists, and the test and retest for the audiovisual in noise condition. In the aforementioned figure, the participants and its SRTs are separated in two groups depending on the training response format (open vs closed). On average, participants improved their SRTs by -1.6 dB SNR on their third training list. The total improvement between the first training list and the test list was -2.9 dB SNR. On the second session, participants retained the same SRT scores in the training as in their last list of the first session. SRTs improved further on the list of the retest session, by -3.8 dB SNR relative to the first training list of the first session. Figure 6 shows that there was a consistent difference of ~1.8 dB SNR between the mean SRTs of the open-set and closed-set lists.

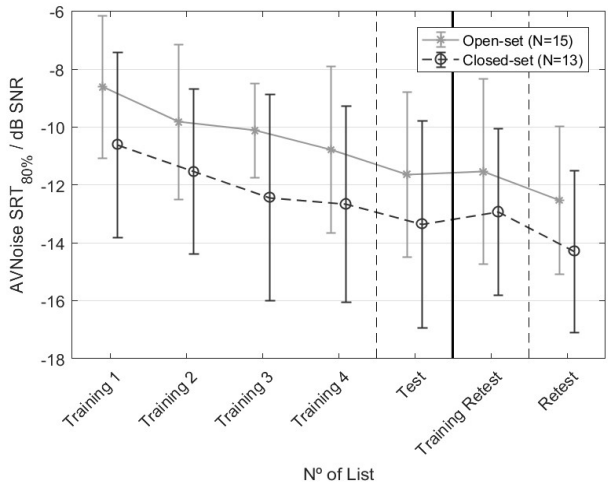

**Figure 6:** Audiovisual training effects. The average and the standard deviation of SRTs over groups are shown. The black dashed line with circles shows the SRTs of the 13 participants that did the training in closed response format. The continuous grey line with whiskers shows the SRTs of the 15 participants that did the training in open response format. It should be noted that, due to the other measurement conditions, there could be up to 4 lists in between the Training 4 and Test lists and between Training Retest and Retest lists.

A repeated-measures ANOVA was performed with response format as the between-subjects factor (open vs. closed) and position within the initial training lists (1st, 2nd, 3rd, and 4th training list) as the within-subjects factor. The dependent variable was the SRT. The sphericity assumption had not been violated according to Mauchly's test ($\chi^2(5) = 2.33$; $p = 0.80$). A significant main effect was found for the within-subjects factor (training list order) ($F(3, 78) = 10.96$; $p < 0.001$). No significant effect was found for the between-subjects factor (response format in the training lists) ($F(1, 26) = 4.17$; $p = 0.052$), although it was close to being significant. No significant interaction was found between the training list's position and the response format ($F(3, 78) = 0.21$; $p = 0.82$). Multiple comparisons with Bonferroni corrections showed that the SRT of the first list was significantly different from the SRTs of the other three. The SRTs of the second, third and fourth list did not differ significantly.

### 3.2 Audio-only and Audiovisual SRTs

Mean SRTs and standard deviations of the lists for the different conditions are shown in Table 3. In the table, test and retest SRTs are grouped together per condition. The average SRT differences between audio-only and audiovisual lists were 5.0 dB SNR for speech in noise and 7.0 dB SPL in quiet. The listeners' PTAs were not significantly correlated with the audio-only in quiet scores (Pearson's $r = 0.15$, $p = 0.11$).

**Table 3.** Mean audio-only and audiovisual SRTs and between-subjects standard deviations in the test and retest sessions (56 scores per cell).

|  | Mean SRT / dB SNR |  | Mean SRT / dB SPL |
|---|---|---|---|
| **AONoiseClosed** | -8.2 (0.9) | **AOQuietClosed** | 17.6 (3.2) |
| **AONoiseOpen** | -8.2 (1.1) | **AOQuietOpen** | 17.8 (2.4) |
| **AVNoiseClosed** | -13.4 (3.2) | **AVQuietClosed** | 10.9 (4.4) |
| **AVNoiseOpen** | -12.9 (3.4) | **AVQuietOpen** | 10.5 (4.6) |

### 3.3 Ceiling Effects

Participants reached SNRs below -20 dB and speech presentation levels below 0 dB SPL (no sound pressure) in the audiovisual conditions. At these levels, there is no contribution of acoustic information to speech reception: the speech detection threshold for the female OLSA is around -16.9 dB SNR in audio-only tests with TSN [22], a threshold that can be theoretically lowered by around -3 dB when adding visual speech [23]. Therefore, below these thresholds (-20 dB SNR and 0 dB SPL), participants used only visual speech in this experiment, i.e., they were speechreading. In consequence, the scores below these thresholds do not represent audiovisual speech perception, but rather visual-only. Figure 7 shows that during the adaptive procedure, participants could reach levels where there was no acoustic contribution.

For the analysis of the data, we decided to limit the values that were below the acoustic speech detection thresholds, as they were not representative of audiovisual speech reception. In total, 18 out of 364 SRTs of audiovisual lists (5%) were modified by limiting them to -20 dB SNR for speech in noise and 0 dB SPL for speech in quiet. We decided to include these scores as they were representing the best speechreading scores. The lists affected had varied conditions: of the 18 lists, 3 were training lists, 5 AudiovisualNoiseOpen, 5 AudiovisualNoiseClosed, 3 AudiovisualQuietOpen, and 2 AudiovisualQuietClosed. Of the 28 participants, 6 were able to go below the speech detection thresholds.

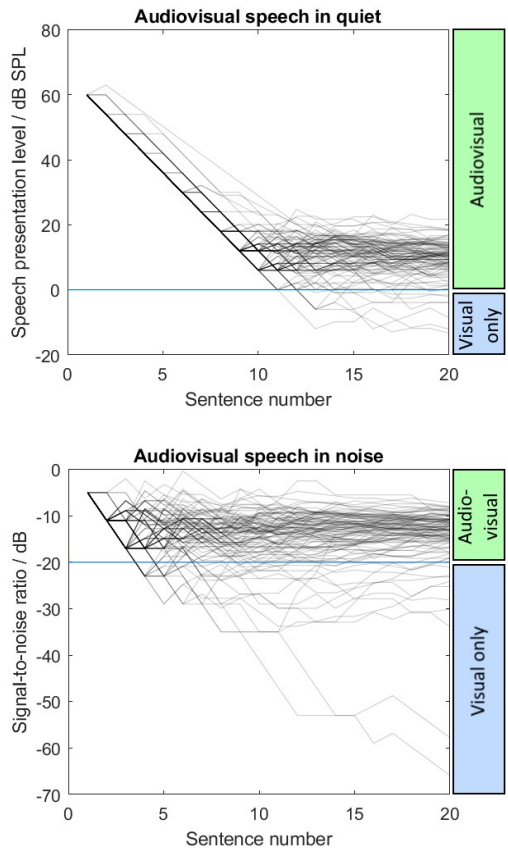

**Figure 7:** Adaptive SNRs and speech presentation levels for AVQuiet (top) and AVNoise (bottom) conditions. The adaptive procedure changed the speech levels to reach 80% intelligibility. Below the blue horizontal line, participants understood speech using only visual cues. Each line shows a single list, adding up to 4 lines per participant in each subfigure.

### 3.4 Speechreading and Audiovisual Benefit

Participants had a wide range of speechreading abilities. The individual VONoiseClosed scores ranged from 0 to 84% intelligibility, had an average of 50% and a standard deviation of 21.4%. Figure 8 shows the distribution of the visual-only scores. There was an average intelligibility improvement of 6.1% in the retest over the test session, although not all participants improved their scores.

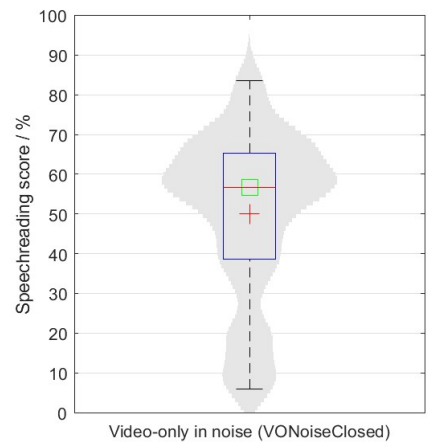

**Figure 8:** Boxplot and distribution of the speechreading scores. In this figure, each participant has a data point: the average word scoring percentage over 40 sentences. The mean and the median are represented as a red cross and a green square, respectively.

Speechreading scores were correlated with the audiovisual benefit (i.e., the SRT difference between audiovisual and audio-only condition). This correlation can be seen in Figure 9, where the visual-only scores are plotted against the individual SRT benefits in different conditions. The Pearson's r correlation scores between the speechreading scores (VONoiseClosed) and the audiovisual benefits were -0.76 (p<0.001) for AVNoiseClosed minus AONoiseClosed, -0.69 (p<0.001) for AVNoiseOpen minus AONoiseOpen, -0.65 (p<0.001) for AVQuietClosed minus AOQuietClosed, and -0.65 (p<0.001) for AVQuietOpen minus AOQuietOpen. Participants that were good speechreaders gained more from having visual information in the audiovisual lists. Whether participants were trained in open-set or closed-set formats did not make any difference for the audiovisual benefit.

Preprint                                                                                                G. Llorach et al.

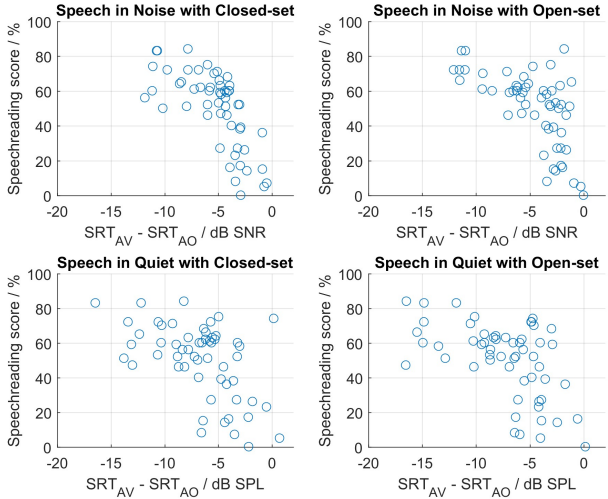

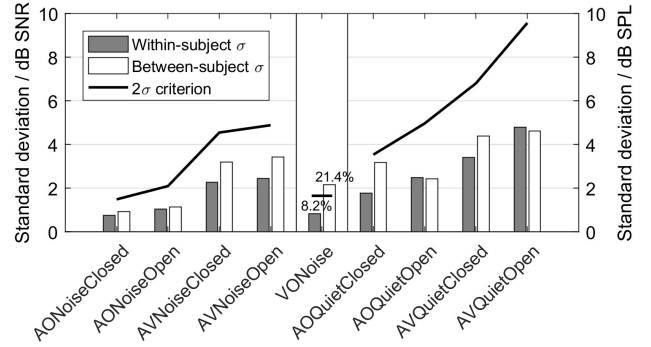

Figure 9: Speechreading scores (VONoiseClosed) shown against the audiovisual benefit of each participant; each participant has two circles per plot for test and retest lists. Top left: audiovisual benefit in noise with closed-set response. This condition was the most similar to the visual-only condition, as both had noise and a closed-set format. Top right: audiovisual benefit in noise with open-set format. Bottom: audiovisual benefit in quiet with closed-set (left) and open-set formats (right).

### 3.5 Test-retest differences

The within-subject and the between-subject standard deviations of the SRTs are shown in Figure 10. The standard deviations of the within-subject differences (test minus retest) are shown as gray bars. The between-subjects standard deviations are shown as white bars. The 2σ criterion is shown as a thick black line. The 2σ criterion represents the threshold where it is possible to differentiate significantly between individuals: if the between-subject standard deviation is higher than the double of the within-subject standard deviation, i.e., the 2σ criterion, it means that it is possible to differentiate significantly between subjects [45]. None of the conditions but the VONoise exceeded the 2σ criterion. The audiovisual conditions had a higher within-subject and between-subject variability in comparison to their respective audio-only conditions.

Figure 10: Within-subject (gray bars) and between-subject (white bars) standard deviations for all conditions. The 2σ criterion is indicated as a thick black line. On the left, STDs of speech in noise conditions expressed in dB SNR; on the middle, STDs of the speechreading scores expressed in percentage; and on the right, STDs of speech in quiet expressed in dB SPL.

## 4 Discussion

### 4.1 Validity of the Video Material

The audio-only and audiovisual scores found were similar to those expected based on the literature. A difference of 3 dB between audio-only and audiovisual scores was reported previously [8], whereas we found a difference of more than 5 dB in the equivalent conditions. This difference could arise from the specific speaker, as some speakers are easier to speechread than others [24], or from language differences [3].

The results of the audiovisual MST were in concordance with the literature, thus validating the video material for measuring speech reception thresholds using lists of 20 sentences. Nevertheless, due to the inherent dubbing asynchronies, the material presented here might not be suitable for investigating fine-grained effects of audiovisual interactions. Audiovisual asynchronies can detriment speech perception [13], but we believe that these asynchronies did not affect severely the results. In another publication [30], the data of this study was analyzed on a word level. Speech intelligibility detriments due to dubbing asynchronies were not looked at, but it was shown that if a word was harder to understand in the audiovisual version it was because this word was hard to speechread. In other words, audiovisual benefits and detriments were explained in [30] by how easy to speechread a word was.

In our study, participants were not specifically asked about audiovisual asynchronies in the audiovisual material during the evaluation, and none reported any temporal artifacts.

### 4.2 Advantages of Optimized and Validated Audio Material

As mentioned in the introduction, one of the advantages of using existing audio material is that it maintains the validity of acoustic speech. For example, van Wanrooij et al. (2019) [8] reported a large variability in intelligibility across words, which probably arose because the word acoustic levels were not balanced and optimized, as is usually done in MSTs. Nevertheless, this does not mean that a non-optimized MST is not usable: MSTs without level adjustments [25] are used in research and can be used to evaluate speech recognition thresholds with almost the same precision.

Another advantage of using existing audio material is that it makes the recording procedure simpler. Limiting the final number of sentences (150) simplifies and speeds up the recording process. Jamaluddin (2016) [6] did not have a final selection of sentences, and so they created all 100,000 possible sentences by re-mixing 100 recorded sentences. During the recording session, they had to ensure that the speaker's head was in the same physical position so that the videos could be cut and blended without artifacts. For this purpose, they had to fabricate a head-resting apparatus to keep the head in the same position. The material required an additional evaluation step to validate the re-mixed recordings, resulting in 600 final sentences.

Another possible solution for creating visual speech, and one that offers more flexibility and control, is animated virtual characters. Ideally, the virtual character's lip-syncing should achieve the same intelligibility scores as the videos of real speakers. Some of the current virtual characters used in audiological research improve speech intelligibility. Schreitmüller et al. (2018) [10] used the German MST with virtual characters: CI and NH participants achieved 37.7% and 12.4% average word scoring in the visual-only condition, respectively. These values are below the scores we found in this study (50%), but they cannot be compared directly because we only considered young normal-hearing listeners. Similarly, Grimm et al. (2019) [39] used the German MST with virtual characters and compared it to the material presented here (AV-OLSAf), but no SRT improvements were found. Devesse et al. (2018) [9] reported an SRT improvement of 1.5 to 2 dB SNR with virtual characters, while we found a 5 dB SNR improvement; the speech material in that study was different from ours and thus cannot be compared directly.

For each research application one has to find the best compromise when creating audiovisual MSTs. For some it might be enough to use synthetic speech and virtual characters with lip-sync, whereas others might need audiovisual synchronous recordings with balanced word acoustic levels. We found that dubbed videos were the most cost-effective solution for the research applications in our laboratory and that it might be a useful technique for others to measure gross audiovisual speech intelligibility.

On a side note, we would like to encourage audiovisual MSTs as a tool for evaluating the lip-syncing animations of virtual characters. Most current research in lip animation and visual speech does not consider human-computer communication and speech understanding in their evaluation procedures [26][27].

### 4.3 Speechreading

The ceiling effects found in the audiovisual MST resulted from the visual speech contribution. These ceiling effects change how the audiovisual MST can be tested. Some participants achieved scores up to 84% just by speechreading. If the audiovisual MST is tested with an adaptive procedure targeting 50% SRT, there will be quite some participants that will be able to speechread half the material without using acoustic information.

Even at 80% SRT, we found few participants that could achieve SNRs were the sentences are not audible anymore. Excluding these data points would have been equivalent to removing the best audiovisual scores. But keeping them as they were would have led to unrealistic audiovisual SNR benefits (some participants reached scores below -60 dB SNR in audiovisual lists). We decided that limiting these values to the level were acoustic information disappears was the best trade-off. Another sensible approach would be to use the median SNR instead of the mean.

These effects could be because the limited set of words in the MST is easy to learn, to differentiate visually, and to speechread. Additionally, because there are only 150 possible sentences, some participants might memorize some of them after several repetitions. However it is rather difficult to memorize the sentences because of their syntactical structure with low context [46]. In sentences for which participants have no previous knowledge of content, one would expect lower speechreading scores, of around 30% [31][38]. Nevertheless, it can be argued that having some expectations about sentence content is probably closer to a real-life conversation.

Another possible factor is that the female speaker was easy to speechread. Additionally, Bench et al. (1995) [24] reported that young female speakers were judged to be easier to speechread than males and older females. We did not make a selection of speakers, as we wanted to have the same person that recorded the audio-only MST. Furthermore, female speakers have been recommended as a compromise between the voice of an adult male and a child [28], so this was a reasonable starting point. Selecting speakers that are more difficult to speechread would probably reduce the ceiling effects.

An interesting alternative to audiovisual MSTs would be to develop a viseme-balanced MST. The audio-only MST is designed to be phonetically balanced, but this does not mean that the visual speech is balanced, as each phoneme does not necessarily correspond to a viseme [29]. Visual cues were previously reported to affect word intelligibility and word error for the AV-OLSAf [30], demonstrating that acoustic speech and visual speech provide different information. Therefore it is possible that the visual speech found in the current MST sentences is not representative of the language tested. Language-specific viseme vocabularies [31] should be developed for this purpose.

That the audiovisual lists were correlated with the speechreading scores was expected [8] [32] [33]. The better a participant was at speechreading, the less acoustic information he or she needed to understand speech. This correlation was present in noise and in quiet conditions; the audiovisual benefit was therefore resilient to the acoustic condition.

### 4.4 Training Effects

An improvement of 2.2 dB SNR between the 1$^{st}$ and the 8$^{th}$ list at 50% speech reception is expected in audio-only MSTs [15]. We found a ~3 dB SNR improvement at 80% speech reception between the first training list and the test list; this additional dB probably arose from the participants learning to speechread the material and becoming familiarized with the speaker [16]. According to the statistical report, the training effect disappeared after one training list. Nevertheless, an average constant improvement was observed. This training effect was not reported in the audiovisual Dutch MST [8] after a familiarization phase with the complete set of words and a training list of 10 audiovisual sentences.

### 4.5 Within- and between- subject variability

In the audio-only speech in noise SRTs, we found little within- and between- subject variability, which was expected, as all participants were young and did not have any hearing disability [35]. Both within- and between-subject variability increased in the audio-only speech in quiet lists, which is expected in quiet conditions [34]. Hearing thresholds and noise-induced hearing loss are usually correlated with speech in quiet scores: the worse the hearing levels, the worse the speech intelligibility in quiet [34]. Nevertheless, we did not find this correlation in our study, probably because the PTAs were all very similar and we did not include hearing-impaired participants.

The speechreading scores were highly individual and diverse in a homogeneous group of participants, which was expected from the literature [6][8]. The test-retest analysis showed that the visual-only lists could differentiate significantly between individuals, meaning that the visual-only MST can assess the speechreading ability of an individual.

The larger between-subject variability found in audiovisual lists can be explained by individual speechreading abilities. If a participant had a high speechreading score, it would be reflected in its audiovisual score. Nevertheless, when looking at the test-retest differences, the within-subject variability in the audiovisual scores did not permit to differentiate between participants significantly. Why could the audiovisual MST not differentiate between participants in the audiovisual modality, given that they all had the same hearing abilities but very different speechreading scores? One possible explanation for the within-subject variability in the audiovisual condition is that the asynchronies of the audiovisual material reduced the test-retest reliability. Another plausible explanation is that the integration between two types of modalities (acoustic and visual) led to a variance that could not be accounted for, assuming that audiovisual integration is an independent modality [47]. Further research should look into the within-subject variability in audiovisual speech perception, as it cannot be derived from this study.

Audiovisual MSTs are particularly relevant for testing severe-to-profound hearing-impaired listeners in the clinic. These listeners cannot perform audio-only intelligibility tests and therefore the audiovisual MST would be useful for investigating whether hearing aid or cochlear implant provision improves their audiovisual speech comprehension. Additionally, the test provides information about the speechreading abilities of an individual. If the individual can speechread well, further recommendations could be provided to the patient for everyday-live situations, such as placing yourself in a position where you can see the mouth of the speakers.

We believe that our material can be used for clinical purposes, when taking into account aforementioned effects: in order to minimize ceiling effects, an 80% SRT is recommended; and at least one or two training list should be used to minimize training effects. Further research should evaluate the AV-OLSAf with hearing-impaired and elderly participants, as some effects are expected: hearing-impaired listeners tend to be better speechreaders [36], and the ability to speechread decreases with age [37]. Furthermore, the influence of the type of noise could change in the audiovisual version and should be investigated [45]. Audiovisual integration needs to be further investigated with specific tests of audiovisual integration and different subject groups, as it has been suggested as an indicator of audiovisual speech intelligibility in noise, especially for those individuals with a hearing loss [48].

## 5 Conclusions

- The method presented here keeps the validity of the original audio material while introducing concordant visual speech. Dubbed video recordings gave similar benefit in terms of gross speech intelligibility measures as naturally synchronous audiovisual recordings, according to literature data, and thus are applicable for our purposes of assessing audiovisual speech intelligibility scores. Other fine-grain effects of audiovisual interaction may not be accessible through the dubbed recordings.

- The audiovisual MST suffers from ceiling effects, which are closely related to the speechreading abilities of the

participant. These effects should be considered when designing experiments for audiovisual perception. High target SRTs such as 80% SRT are recommended instead of 50% SRT in adaptive procedures.

- Audiovisual stimuli gave an SRT benefit of 5 dB SNR in test-specific noise and 7 dB SPL in quiet in comparison to audio-only stimuli for young, normal-hearing participants. Reference values for 80% SRT found in this study were -13.2 dB SNR for audiovisual speech in noise and 10.7 dB SPL for audiovisual speech in quiet.

- At least one training list should be completed in order to avoid statistically significant training effects. These effects may continue after a certain number of training lists. It is therefore recommended that two training lists are used to evaluate an audiovisual condition.

- Audiovisual SRTs correlated with speechreading abilities. The better participants could speechread, the more they benefited in the audiovisual conditions.

- The visual-only MST can be used to differentiate between the speechreading abilities of young normal-hearing individuals. Due to the variability in the audiovisual SRTs, we recommend including a visual-only condition when assessing audiovisual speech perception with the AV-OLSAf.

## ACKNOWLEDGMENTS

This work received funding from the EU's H2020 research and innovation program under the MSCA GA 675324 (ENRICH), from the Deutsche Forschungsgemeinschaft (DFG, German Research Foundation) – project number 352015383 (SFB 1330 B1 and C4) and from the European Regional Development Fund – Project "Innovation network for integrated, binaural hearing system technology (VIBHear)". We would like to thank the Media Technology and Production of the CvO University of Oldenburg for helping out with the recordings. Special thanks to Anja Gieseler for giving feedback on the evaluation procedures and the manuscript and to Bernd T. Meyer for counseling on the video selection metric.